\documentclass{pasj00}
\Received{$\langle$2015 May 16$\rangle$}
\Accepted{$\langle$2015 June 12$\rangle$}
\Published{$\langle$$\rangle$}

\usepackage{epsfig}
\usepackage{subfigure}
\usepackage{graphicx}
\usepackage{rotating}
\usepackage{natbib}
\usepackage{booktabs}
\usepackage{bm}

\usepackage{threeparttable}

\makeatletter
\newcommand\footnoteref[1]{\protected@xdef\@thefnmark{\ref{#1}}\@footnotemark}
\makeatother

\begin{document}

\title{Revisiting a gravity-darkened and precessing planetary system PTFO 8-8695: spin-orbit non-synchronous case}
\author{Shoya \textsc{Kamiaka}\altaffilmark{1},
Kento \textsc{Masuda}\altaffilmark{1},
Yuxin \textsc{Xue}\altaffilmark{1},
Yasushi \textsc{Suto}\altaffilmark{1,2},
Tsubasa \textsc{Nishioka}\altaffilmark{3},
Risa \textsc{Murakami}\altaffilmark{3},
Koichiro \textsc{Inayama}\altaffilmark{3},
Madoka \textsc{Saitoh}\altaffilmark{3},
Michisuke \textsc{Tanaka}\altaffilmark{3},
and Atsunori \textsc{Yonehara}\altaffilmark{3}}
\altaffiltext{1}{Department of Physics, The University of Tokyo, Tokyo 113-0033, Japan}
\altaffiltext{2}{Research Center for the Early Universe, The University of Tokyo, Tokyo 113-0033, Japan}
\altaffiltext{3}{Department of Physics, Kyoto Sangyo University, Kyoto 603-8555, Japan}

\email{kamiaka@utap.phys.s.u-tokyo.ac.jp}
 

\maketitle

\begin{abstract}
We reanalyse the time-variable lightcurves of the transiting planetary system PTFO 8-8695, in which a planet of 3 to 4 Jupiter mass orbits around a rapidly rotating pre-main-sequence star.
Both the planetary orbital period $P_{\rm{orb}}$ of 0.448 days and the stellar spin period $P_{\rm{s}}$ less than 0.671 days are unusually short, which makes PTFO 8-8695 an ideal system to check the model of gravity darkening and nodal precession.
While the previous analysis of PTFO 8-8695 assumed that the stellar spin and planetary orbital periods are the same, we extend the analysis by discarding the spin-orbit synchronous condition, and find three different classes of solutions roughly corresponding to the nodal precession periods of $199\pm16$, $475\pm21$, and $827\pm53$ days that reproduce the transit lightcurves observed in 2009 and 2010.
We compare the predicted lightcurves of the three solutions against the photometry data of a few percent accuracy obtained at Koyama Astronomical Observatory in 2014 and 2015, and find that the solution with the precession period of $199\pm16$ days is preferred even though preliminary.
Future prospect and implications to other transiting systems are briefly discussed.
\end{abstract}

\section{Introduction \label{sec:introduction}}

PTFO 8-8695 is a pre-main-sequence star (weak-line T-Tauri star) located in the Orion OB1a star forming region, first discovered by \citet{2005AJ....129..907B}.
Subsequent photometric transit and spectroscopic radial velocity measurements revealed that PTFO 8-8695 harbors one close-in planet (PTFO 8-8695 b) with planetary mass $M_{\rm{p}} < 5.5\:M_{\rm{J}}$ \citep{2012ApJ...755...42V}.
This is the first transiting exoplanet candidate around a pre-main-sequence star. 

\citet{2012ApJ...755...42V} discovered that PTFO 8-8695 in 2009 and 2010 exhibited very different transit lightcurves.
\citet{2013ApJ...774...53B} (hereafter, B13) found that the peculiar time-variation can be explained by the gravity darkening of the central star and the spin-orbit nodal precession of the system.
Indeed they are successful in fitting simultaneously the lightcurves in 2009 and 2010, and estimated the system parameters including stellar and planetary radii ($R_{\rm{s}}{\sim}1.0\:R_{\odot}, R_{\rm{p}}{\sim}1.7\:R_{\rm{J}}$), planetary mass ($M_{\rm{p}}=3.0$ or $3.6\:M_{\rm{J}}$), and spin-orbit angle ($\phi{\sim}70^{\circ}$; the angle between stellar spin and planetary orbital axes).

In doing so, they assumed that the stellar spin and planetary orbital motion are synchronised for simplicity.
Given the large spin-orbit misalignment estimated for the PTFO 8-8695 system, however, the synchronous assumption may not be justified from the dynamical point-of-view.
Therefore we reanalyse the PTFO 8-8695 transit photometry without adopting the spin-orbit synchronous assumption.
As expected, the permitted parameter space is increased, and we discuss a feasibility to break the degeneracy among different solutions with future observations. 

The rest of the present paper is organized as follows.  Section 2 briefly summarizes the previous work on the PTFO 8-8695 system.
We adopt a model of gravity darkening and spin-orbit nodal precession of B13, which is described in Section 3. 
Our data analysis and parameter fitting procedures are presented in \S 4.1 and 4.2, respectively, and the resulting constraint on the system parameters from 2009 and 2010 data is shown in \S 4.3.
We also discuss how the long-term photometric monitoring of the system is useful in further constraining the parameter space.
A preliminary analysis with the observation performed in 2014-2015 at Koyama Astronomical Observatory (KAO) is presented in \S 4.4.
We discuss the long-term dynamical stability of the system in Section 5.
Final section is devoted to the summary of the paper and the further discussion.

\section{Exoplanetary system PTFO 8-8695 \label{sec:background}}

The transits of PTFO 8-8695 b were observed at Palomar Observatory for 2009 December 1 - 2010 January 15 (hereafter, referred to as 2009 observation), and 2010 December 8 - 17 (2010 observation) as part of the Palomar Transient Factory (PTF) that ran for 2009 - 2012.
The 2009 and 2010 observations comprise 11 and 6 reliable transit lightcurves, respectively, with the transit period of 0.448413 days \citep{2012ApJ...755...42V}.
As already mentioned in Introduction, the lightcurves exhibit a significant time variation, and B13 solved this puzzle by taking into account both the rapid rotation of the star and short transit period of the planet.

Spectroscopic measurement of PTFO 8-8695 implies that the host star is a rapid rotator with the spin period of $P_{\rm{s}}<$ 0.671 days \citep{2012ApJ...755...42V}.
This is consistent with the observational evidence that T-Tauri stars (especially, weak-line T-Tauri stars like PTFO 8-8695) tend to rotate more rapidly than main sequence stars \citep{1993AJ....106..372E}.
Because of the rapid rotation, the equator of PTFO 8-8695 would be significantly expanded relative to its polar radius.

Because of the stellar surface distortion, PTFO 8-8695 should exhibit significant $\emph{gravity darkening}$ (\citealt{1924MNRAS..84..665V}, \citealt{1967ZA.....65...89L}) and the nodal precession of stellar spin and planetary orbital axes.
B13 took account of both effects properly, and are successful in reproducing both 2009 and 2010 lightcurves either for two different stellar masses, $M_{\rm{s}}$ = 0.34 and 0.44 $M_{\odot}$ \citep{2005AJ....129..907B}.

The nodal precession period depends sensitively on both the stellar spin and planetary orbital periods, $P_{\rm{s}}$ and $P_{\rm{orb}}$.
While the latter is well-determined from the transit period ($0.448$ days), only the upper limit is obtained observationally for the former, $P_{\rm{s}} <$ 0.671 days.
Since \citet{2012ApJ...755...42V} concluded that the stellar spin and planetary orbital motion are most likely synchronized ($P_{\rm{s}} = P_{\rm{orb}}$), B13 adopted that assumption for simplicity throughout their analysis.

Theoretically speaking, however, the validity of this assumption is not clear.
One of the most probable mechanisms for the spin-orbit synchronization is the tidal effect between the star and planet.
In this case, the spin-orbit synchronization and the spin-orbit alignment occur simultaneously.
The equilibrium tidal model, for instance, predicts that the effective timescales for the two processes are almost the same (\citealt{2012MNRAS.423..486L}, \citealt{2013ApJ...769L..10R}).
Nevertheless the result of B13 indicates the significant spin-orbit misalignment ($\phi{\sim}70^{\circ}$), and indeed they admitted that the truly synchronous rotation might be difficult to be achieved under such a misaligned state.
This is why we attempt in the present paper to reanalyse the PTFO 8-8695 system, to see how the estimated parameters are sensitive to the synchronous condition.

\section{Basic equations \label{sec:equation}}

This section summarizes the basic equations that we use in the later analysis, largely following B13.
Our model employs 16 parameters in total, and for clarity we list them in Table \ref{parameters}.

\begin{table*}[!htbp]
\centering
\caption{Parameters for the analysis.}
\begin{threeparttable}
\begin{tabular}{cccc}
\toprule
parameter & symbol & fixed value & note \\
\midrule
Mean stellar density & $\rho_{\rm{s}}$ & - & a \\
Stellar effective temperature at the pole & $T_{\rm{pol}}$ & 3470 K & b \\
Stellar rotation period & $P_{\rm{s}}$ & - & \\
Time of inferior conjunction & $t_{\rm{c}}$ & 2455543.9402 HJD & c, d \\
Stellar inclination at $t_{c}$ & $i_{\rm{s}}$ & - & e \\
Stellar moment of inertia coefficient & $C$ & 0.059 & f \\
limb-darkening parameter & $c_{1}=u_{1}+u_{2}$ & 0.735 & f \\
limb-darkening parameter & $c_{2}=u_{1}-u_{2}$ & 0.0 & \\
Gravity-darkening parameter & $\beta$ & 0.25 & f \\
Planet-to-star mass ratio & $M_{\rm{p}}/M_{\rm{s}}$ & - & \\
Planet-to-star radius ratio & $R_{\rm{p}}/R_{\rm{s}}$ & - & a \\
Planetary orbital period & $P_{\rm{orb}}$ & 0.448413 days & d \\
Orbital eccentricity & $e\cos{\omega}$ & 0 & g \\
Orbital eccentricity & $e\sin{\omega}$ & 0 & g \\
Planetary orbital inclination at $t_{c}$ & $i_{\rm{orb}}$ & - & e \\
Longitude of the ascending node at $t_{c}$ & $\Omega$ & - & e \\
\bottomrule
\end{tabular}
\begin{tablenotes}\footnotesize
\item[a] specified with stellar $\emph{equatorial}$ radius $R_{\rm{s},\rm{eq}}$; $M_{\rm{s}}/\frac{4}{3}{\pi}R_{\rm{s},\rm{eq}}^3$ or $R_{\rm{p}}/R_{\rm{s},\rm{eq}}$.
\item[b] \citet{2005AJ....129..907B} reported 3470 K as the stellar effective temperature, but we use that value as polar temperature assuming that their difference in the analysis is negligible.
\item[c] an epoch when $\omega + f = \pi/2$.
\item[d] \citet{2012ApJ...755...42V}.
\item[e] \citet{2013ApJ...774...53B} adopt different notations for these three angular parameters (stellar obliquity $\psi$, planetary orbital inclination $i$, projected spin-orbit angle $\lambda$) from ours, and they are related as $\psi=i_{\rm{s}}-\pi/2$, $i=\pi-i_{\rm{orb}}$, and $\lambda=\pi-\Omega$.
\item[f] \citet{2013ApJ...774...53B}.
\item[g] We assume a circular orbit following \citet{2013ApJ...774...53B}.
This assumption is supported by an equilibrium tidal theory, which predicts that a close-in planet acquires the circular orbit on a much shorter time scale than those for spin-orbit synchronization or alignment.
\end{tablenotes}
\end{threeparttable}
\label{parameters}
\end{table*}%

\subsection{Configuration of the system \label{subsec:configuration}}

\begin{figure}[!h]
\centering
\includegraphics[width=85mm]{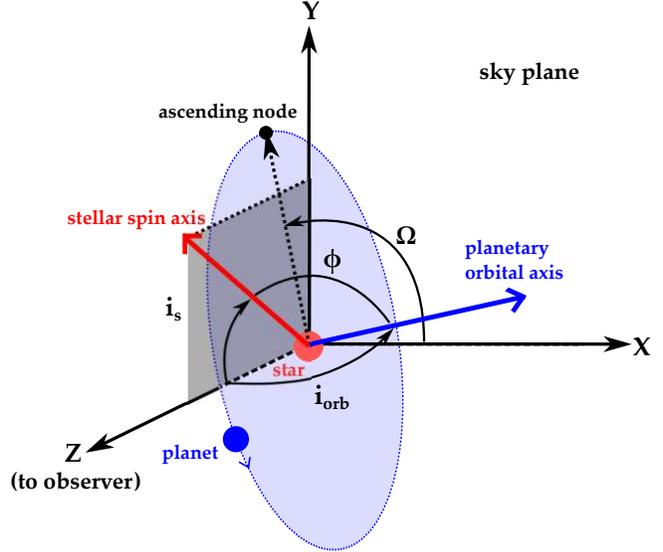}
\caption{Schematic illustration for spin-orbit angle $\phi$ as a function of stellar inclination $i_{\rm{s}}$, planetary orbital inclination $i_{\rm{orb}}$, and longitude of the ascending node $\Omega$.}
\label{angles}
\end{figure}

Figure \ref{angles} defines the parameters specifying the geometric configuration of the system.
We define the Cartesian coordinates in such a way that the star is located at the origin and the $x-y$ plane coincides with the sky plane with the positive $z$-direction pointing towards the observer.
The stellar spin vector projected onto the sky plane is defined as the positive $y$-direction, and the $x$-direction is set so as to form a right-handed triad.

The stellar inclination and planetary orbital inclination, $i_{\rm{s}}$ and $i_{\rm{orb}}$, respectively, are measured from the line-of-sight pointing to the observer (positive $z$-axis).
Longitude of the ascending node, $\Omega$, is measured from the positive $x$-axis to the direction towards the ascending node counterclockwise.

In what follows, $\bm{S}$ and $\bm{L}$ denote the stellar spin and planetary orbital angular momentum vectors, respectively.
We define the total angular momentum vector as $\bm{J} = \bm{S} + \bm{L}$.
Their unit vectors and norms are denoted as $\bm{s}$, $\bm{l}$, and $\bm{j}$, and $S$, $L$, and $J$, respectively.

In terms of the angles defined in Figure \ref{angles}, $\bm{s}$ and $\bm{l}$ are written as
\begin{eqnarray}
\bm{s}
=
\left(
\begin{array}{c}
0 \\
\sin{i_{\rm{s}}} \\
\cos{i_{\rm{s}}}
\end{array}
\right),
\quad
\bm{l}
=
\left(
\begin{array}{c}
\sin{i_{\rm{orb}}}\sin{\Omega} \\
-\sin{i_{\rm{orb}}}\cos{\Omega} \\
\cos{i_{\rm{orb}}}
\end{array}
\right).
\end{eqnarray}
Also the spin-orbit angle $\phi$ is written as
\begin{eqnarray}
\phi
&=&
\cos^{-1}(\bm{s}{\cdot}\bm{l})\nonumber
\\
&=&
\cos^{-1}(-\sin{i_{\rm{s}}}\sin{i_{\rm{orb}}}\cos{\Omega} +\cos{i_{\rm{s}}}\cos{i_{\rm{orb}}}).
\label{phi}
\end{eqnarray}

\subsection{Flux profile of the central star \label{subsec:flux}}

We compute the stellar flux by numerically integrating the stellar intensity profile $B_{\lambda}(T){\times}I(\mu)$ throughout the visible stellar disk, where $B_{\lambda}(T)$ is the Planck function at the wavelength of $\lambda$, and $I(\mu)$ is the limb-darkening function.
The following analysis is performed at the monochromatic wavelength of $\lambda=0.658\:{\mu}$m (see section 4).
We adopt quadratic limb-darkening law in our analysis:
\begin{eqnarray}
I(\mu) = 1 - u_{1}(1-\mu) - u_{2}(1-\mu)^2,
\end{eqnarray}
where $c_{1}=u_{1}+u_{2}$ and $c_{2}=u_{1}-u_{2}$ are quadratic limb-darkening coefficients, and $\mu$ denotes the cosine of the angle between line of sight and the stellar surface normal.
The effective temperature $T$ at each point of the stellar disk is estimated with von Zeipel's theorem \citep{1924MNRAS..84..665V}, which says that the temperature distribution on the stellar surface induced by gravity darkening is specified by the stellar surface gravity $g$:
\begin{eqnarray}
\frac{T}{T_{\rm{pol}}}
= \left(\frac{g}{g_{\rm{pol}}}\right)^{\beta},
\label{zeipel}
\end{eqnarray}
where $T_{\rm{pol}}$ and $g_{\rm{pol}}$ are the effective temperature and surface gravity at the pole, and $\beta$ is the gravity-darkening parameter.
See \citet{2009ApJ...705..683B}, \citet{2011ApJS..197...10B}, and \citet{2015ApJ...805...28M} for more details of our model.

\subsection{Spin-orbit nodal precession
\label{subsec:nodal_precession}}

We solve the orbit-averaged equations of motion for the two rigidly-rotating bodies \citep{2009Icar..201..750B}:
\begin{eqnarray}
\label{eq:s}
\dot{\bm{s}} &=& -\frac{p}{S}\bm{l}{\times}\bm{s},\\
\label{eq:l}
\dot{\bm{l}} &=& -\frac{p}{L}\bm{s}{\times}\bm{l},
\end{eqnarray}
where
\begin{eqnarray}
p&=&\frac{3}{2}\frac{GM_{\rm{p}}}{a^3(1-e^2)^{3/2}}J_{\rm{2}}M_{\rm{s}}R_{\rm{s}}^2(\bm{s}{\cdot}\bm{l})\nonumber\\
&=&\frac{3}{2}n^2\frac{M_{\rm{s}}M_{\rm{p}}}{M_{\rm{s}}+M_{\rm{p}}}\frac{1}{(1-e^2)^{3/2}}J_{\rm{2}}R_{\rm{s}}^2\cos{\phi}\nonumber\\
&=&n\cos{\phi}\frac{3}{2}J_{\rm{2}}\left(\frac{R_{\rm{s}}}{a}\right)^2\frac{1}{(1-e^2)^2}{\times}L\nonumber\\
&{\equiv}&-\dot{\Omega}_{\rm{p}}L.
\end{eqnarray}
Here $a$, $n=\sqrt{G(M_{\rm{s}}+M_{\rm{p}})/a^3}=2{\pi}/P_{\rm{orb}}$, and $J_{2}$ denote orbital semi major axis, orbital mean motion, and the second-order gravitational coefficient of the central star, respectively.
In this formulation the planet is considered as a point mass (i.e., planetary spin is neglected) and higher-order gravitational moments are neglected.

Equations (\ref{eq:s}) and (\ref{eq:l}) are rewritten in terms of $\bm{J}=\bm{S}+\bm{L}=S\bm{s}+L\bm{l}$:
\begin{eqnarray}
\dot{\bm{s}} &=& -\frac{p}{SL}(\bm{J}-S\bm{s}){\times}\bm{s} 
= -\frac{pJ}{SL}\bm{j}{\times}\bm{s}
= \dot{\Omega}_{\rm{p}}\frac{J}{S}\bm{j}{\times}\bm{s},\\
\dot{\bm{l}} &=& -\frac{p}{LS}(\bm{J}-L\bm{l}){\times}\bm{l}
= -\frac{pJ}{LS}\bm{j}{\times}\bm{l} 
= \dot{\Omega}_{\rm{p}}\frac{J}{S}\bm{j}{\times}\bm{l}.
\end{eqnarray}
Therefore the angular frequency with which $\bm{s}$ and $\bm{l}$ precess around the total angular momentum $\bm{j}$ is given by
\begin{eqnarray}
\dot{\Omega} &=& \dot{\Omega}_{\rm{p}}\frac{J}{S} =\dot{\Omega}_{\rm{p}}\sqrt{\left(\frac{L}{S}\right)^2+2\left(\frac{L}{S}\right)\cos{\phi}+1}\nonumber\\
&=& \dot{\Omega}_{\rm{p}}\sqrt{\left(\frac{L}{S}+\cos{\phi}\right)^2+\sin^{2}{\phi}}.
\end{eqnarray}

For a circular orbit ($e=0$), $\dot{\Omega}$ reduces to
\begin{eqnarray}
\dot{\Omega} = -n\cos{\phi}\frac{3}{2}J_{2}\left(\frac{R_{\rm{s}}}{a}\right)^2\sqrt{\left(\frac{L}{S}+\cos{\phi}\right)^2+\sin^{2}{\phi}},
\label{precession_frequency}
\end{eqnarray}
with
\begin{eqnarray}
\frac{L}{S} = \frac{1}{C}\frac{M_{\rm{p}}}{M_{\rm{s}}}\frac{n}{\omega_{\rm{s}}}\left(\frac{a}{R_{\rm{s}}}\right)^2,
\end{eqnarray}
where $C$ is the moment of inertia coefficient of the central star and $\omega_{\rm{s}}=2{\pi}/P_{\rm{s}}$ is the stellar spin frequency \citep{2015ApJ...805...28M}.
It can be shown that $\dot{\Omega}$ of equation (\ref{precession_frequency}) is identical to the precession frequency in B13; see their equations (6) or (8).

The standard stellar evolution theory suggests that T-Tauri stars do not yet develop an internal radiative core.
Thus PTFO 8-8695 is expected to be fully convective, and $J_{2}$ in this case is given by \citep{1999ssd..book.....M}
\begin{eqnarray}
J_{2}
=Cf
= C\frac{\omega_{\rm{s}}^2R_{\rm{s}}^3}{2GM_{\rm{s}}} 
=\frac{C}{2}\left(\frac{\omega_{\rm{s}}}{n}\right)^2\left(\frac{R_{\rm{s}}}{a}\right)^3,
\end{eqnarray}
where
\begin{eqnarray}
f &=& \frac{R_{\rm{s},\rm{eq}}-R_{\rm{s},\rm{pol}}}{R_{\rm{s},\rm{eq}}} 
\approx \frac{\omega_{\rm{s}}^2R_{\rm{s},\rm{eq}}^3}{2GM_{\rm{s}}}
\end{eqnarray}
is the oblateness of the central star, and $R_{\rm{s},\rm{eq}}$ and $R_{\rm{s},\rm{pol}}$ are stellar equatorial and polar radii, respectively.

Since Kepler's third law is written as
\begin{eqnarray}
\frac{R_{\rm{s}}}{a}
=\left(\frac{3n^2}{4{\pi}G\rho_{\rm{s}}}\right)^{1/3},
\end{eqnarray}
$\dot{\Omega}$ in equation (\ref{precession_frequency}) is determined by the six parameters $(n,\:\phi,\:\omega_{\rm{s}},\:\rho_{\rm{s}},\:C,\:M_{\rm{p}}/M_{\rm{s}})$.
Among these, the first four are constrained from the transit shape, and we adopt $C=0.059$, the value for the Sun, following B13.
Thus $M_{\rm{p}}/M_{\rm{s}}$ is the only parameter that is additionally constrained from the precession analysis.

Note that since the equations of motion depend on $\rho_{\rm{s}}$ alone, we cannot evaluate the stellar mass ($M_{\rm{s}}$) and radius ($R_{\rm{s}}$) separately.
We also emphasize here that the current analysis does not fix the absolute mass or radius ($M_{\rm{s}}, M_{\rm{p}}, R_{\rm{s}}, R_{\rm{p}}$), but their ratios ($M_{\rm{p}}/M_{\rm{s}}, R_{\rm{p}}/R_{\rm{s}}$) only.

\section{Data analysis: methods and results \label{sec:analysis}}

\subsection{Data reduction \label{subsec:reduction}}

The photometric data analysed by B13 are kindly provided to us by Julian van Eyken, which include 11 reliable transits in 2009 and 6 in 2010 observed at 0.658 ${\mu}$m.
The effects of stellar activity, long-term periodical fluctuations (for instance, due to stellar spots) and background noise are already removed and whitened.  

We repeat the data reduction ourselves again following B13:
\begin{enumerate}
\item Phase-fold.
All transit lightcurves are stacked with the orbital period of $0.448413$ days. 
\item Clipping.
During orbital period of $0.448413$ days, the transit actually continues at best for ${\sim}0.2$ days.
Thus a major part of the phase-folded lightcurve corresponds to out-of-transit phase.
In order to reduce the number of the data points and subsequent computational cost, we focus on $\pm$ 0.1 days around the transit center.
\item Binning.
All data points are combined into one-minute bins.
\item Positioning the 2009 and 2010 lightcurves.
B13 found that the transit centers for 2009 and 2010 data are 30861700 and 60848300 s measured from 2009 January 1 UTC.
Therefore we position 2009 and 2010 lightcurves at those epochs.
The mutual separation between the 2009 and 2010 lightcurves corresponds to 774 transits.
\end{enumerate}

\subsection{Parameter fitting procedure \label{subsec:strategy}}

Next we perform the parameter fitting as follows:
\begin{enumerate}
\item Calculate $\dot{\Omega}$ assuming a set of values for the parameters listed in Table \ref{parameters}.
\item Calculate $\bm{s}$ and $\bm{l}$ at the observed epochs in 2009 and 2010 by extrapolating those at $t_{c}$ through the precession model prediction.
\item Evaluate angular parameters ($i_{\rm{s}}$, $i_{\rm{orb}}$, $\Omega$) at the observed epochs in 2009 and 2010.
\item Compute the model lightcurves at those epochs using a routine developed by \citet{2015ApJ...805...28M}, and evaluate $\chi^2 = \sum_{i}(\rm{flux}_{\rm{data},\it{i}} - \rm{flux}_{\rm{model},\it{i}})^2/\sigma_{\rm{data},\it{i}}^2$ between model and phase-folded lightcurves, where $i$ indicates the variables evaluated at time $t_{i}$ in 2009 and 2010.
\end{enumerate}
The above procedures are iterated so as to find the set of parameters that minimizes $\chi^2$ with the Levenberg-Marquardt algorithm \citep{2009ASPC..411..251M}.
In what follows, we discuss in terms of the reduced chi-square $\chi_{r}^2$ (i.e., $\chi^2$ normalized by the degree of freedom of the analysis of 565).

We see how the minimum $\chi_{r}^2$ value differs on $P_{\rm{s}}-M_{\rm{p}}/M_{\rm{s}}$ plane in our analysis.
This strategy makes it easer to compare the results with and without the spin-orbit synchronous condition.
The result of B13 corresponds to $P_{\rm{s}} = 0.44841$ days line.
Also future constraints on $M_{\rm{p}}$ from radial velocity measurements will be easily incorporated in this plane.

We search for the best-fit parameters over the following regions: $0.12<\rho_{\rm{s}}\:(\rm{g/cm^3})<1.21$, $0<R_{\rm{p}}/R_{\rm{s}}<0.5$, $\pi/2<i_{\rm{s}}<\pi$, $0<i_{\rm{orb}}<\pi$, and $0<\Omega<2\pi$.
The lower and upper limits of $\rho_{\rm{s}}$ are adopted from B13.
Note that the symmetry in the gravity-darkened lightcurves introduces the degeneracy on the angular configuration of the system \citep{2011ApJS..197...10B}; $(i_{\rm{s}}, i_{\rm{orb}}, \Omega)\Leftrightarrow(\pi-i_{\rm{s}},\pi-i_{\rm{orb}},-\Omega)$.
While $i_{\rm{s}}$ is originally defined between 0 and $\pi$, therefore, we confine $i_{\rm{s}}$ between $\pi/2$ and $\pi$.
We abandon those solutions that violate the stellar break-up condition beyond which the star is not stable gravitationally:
\begin{eqnarray}
P_{\rm{s}} > 2\pi\sqrt{\frac{3}{4{\pi}G\rho_{\rm{s}}}}.
\end{eqnarray}

\subsection{Results \label{subsec:results}}

\begin{figure}[!h]
\centering
\subfigure{
\includegraphics[width=90mm]{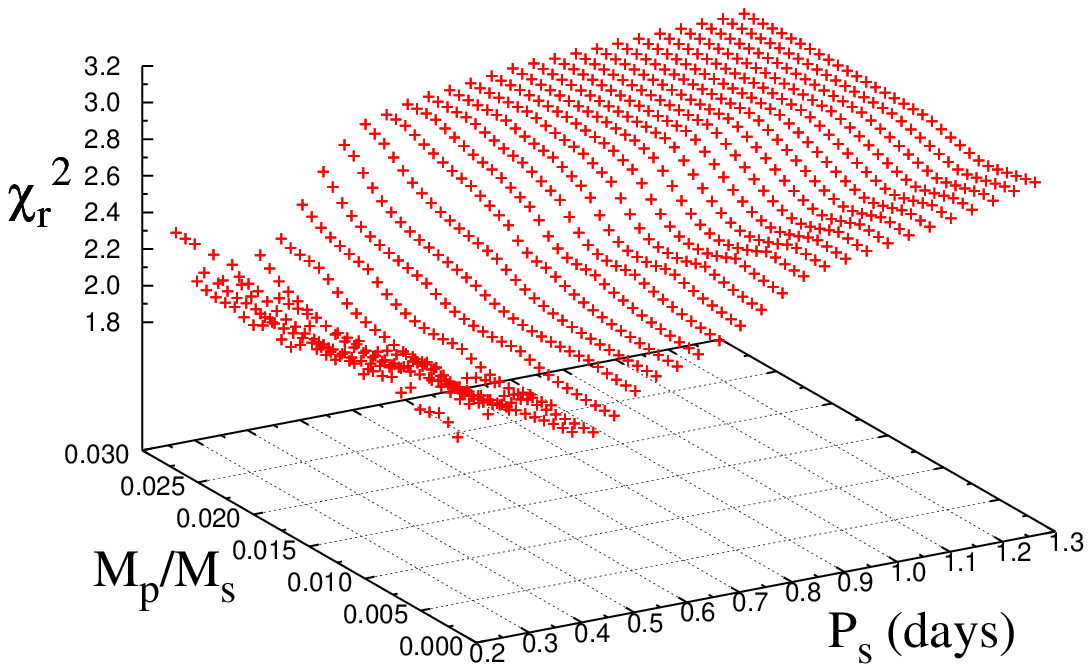}
}%
\\
\vspace{-15mm}
\subfigure{
\includegraphics[width=85mm]{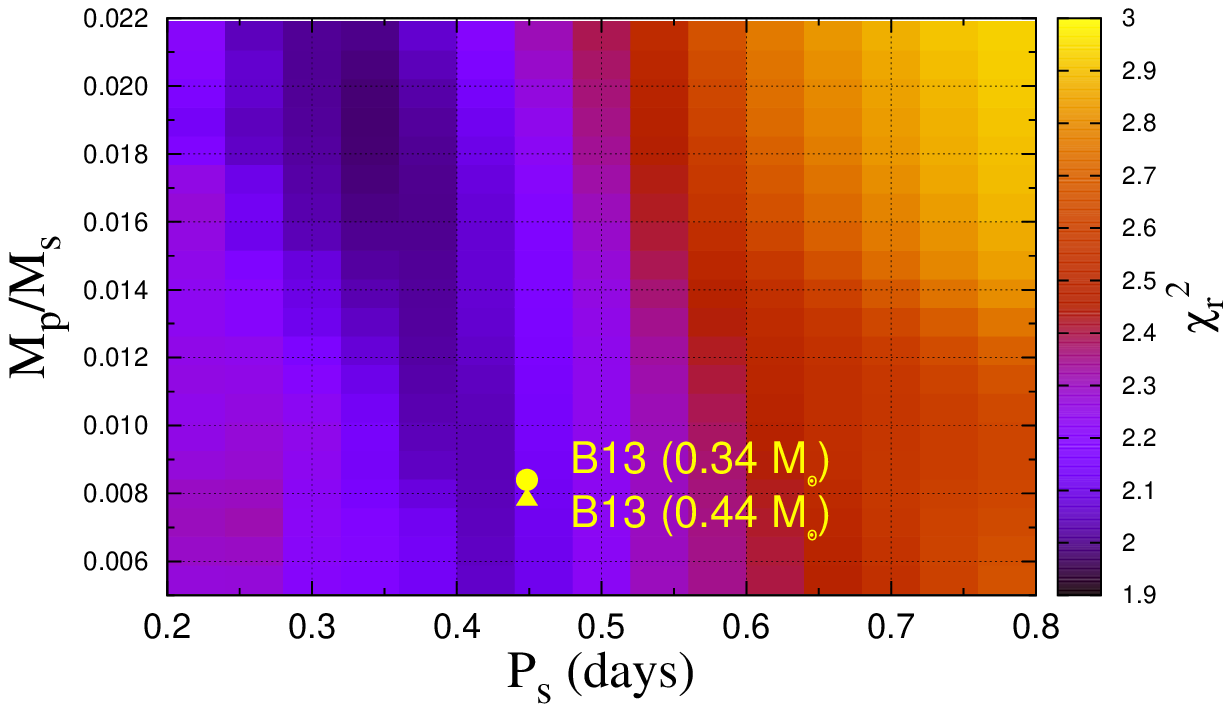}
}%
\vspace{0mm}
\caption{Minimum $\chi_{r}^2$ values for various values of $P_{\rm{s}}$ and $M_{\rm{p}}/M_{\rm{s}}$.
The upper panel shows the three dimensional distribution of minimum $\chi_{r}^2$, while lower panel is the corresponding color contour map.
The results of \cite{2013ApJ...774...53B} are those along the line of $P_{\rm{s}} = 0.44841$ days.
Best solutions found by \cite{2013ApJ...774...53B} are denoted as a yellow circle for $M_{\rm{s}}$ = 0.34$M_{\odot}$ and a yellow triangle for $M_{\rm{s}}$ = 0.44 $M_{\odot}$.}
\label{contour}
\end{figure}

Figure \ref{contour} shows the minimum $\chi_{r}^2$ values on the $P_{\rm{s}}$-$M_{\rm{p}}/M_{\rm{s}}$ plane.
The upper panel shows the three dimensional view, while the lower panel illustrates the corresponding plot as a color-coded contour map.
The analysis of B13 was performed along the line of $P_{\rm{s}} = 0.44841$ days, and their best solutions are plotted as a yellow circle for $M_{\rm{s}}$ = 0.34 $M_{\odot}$ and a yellow triangle for $M_{\rm{s}}$ = 0.44 $M_{\odot}$. 

Figure \ref{contour} suggests that slightly better, or at least equally good, fits are obtained outside the synchronous condition ($P_{\rm{s}} = 0.448413$ days), i.e., in the more rapid stellar spin and massive planet regime.
Therefore, the synchronous condition over-constrains the system parameters, at least unless we have more precise observational data for the stellar spin period, $P_{\rm{s}}$.

\begin{figure}[!h]
\centering
\subfigure{
\includegraphics[width=88mm]{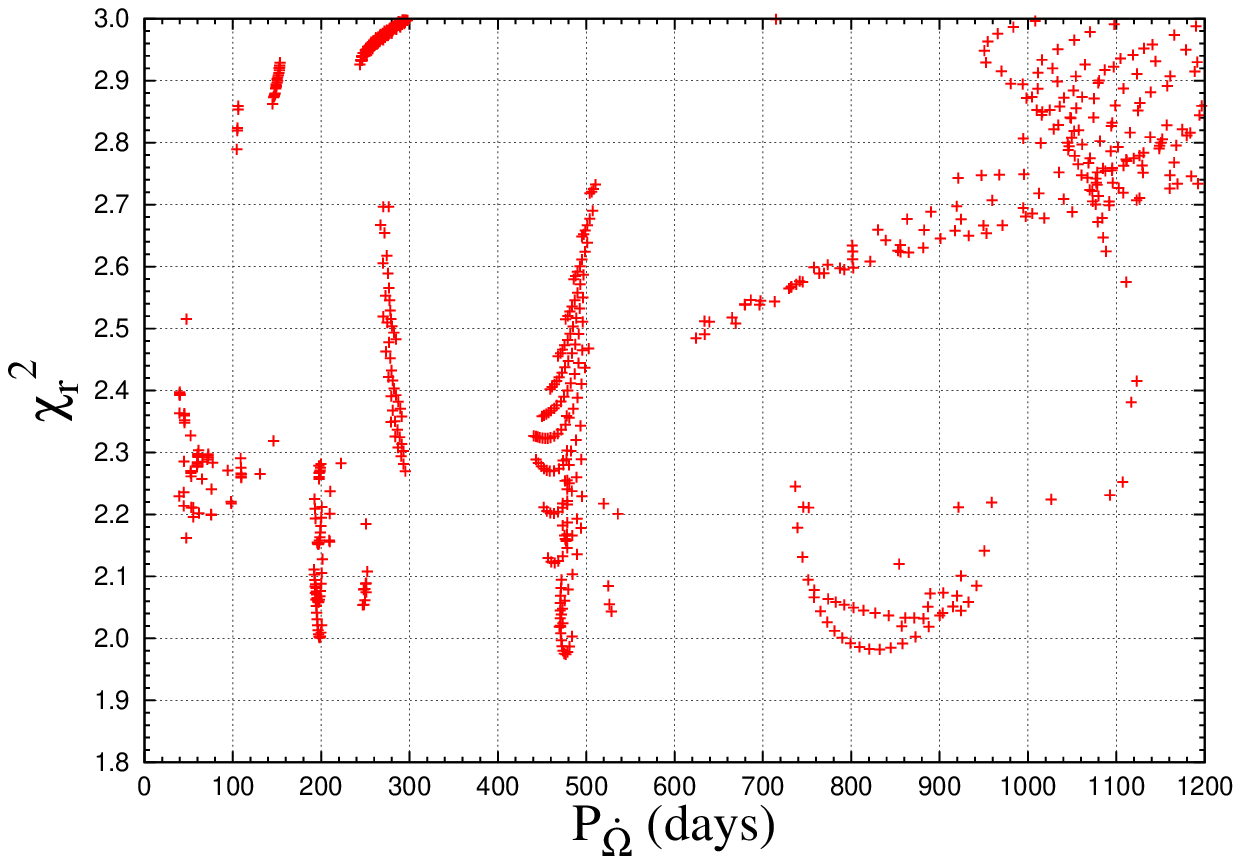}
}%
\\
\vspace{-2.5mm}
\subfigure{
\includegraphics[width=85mm]{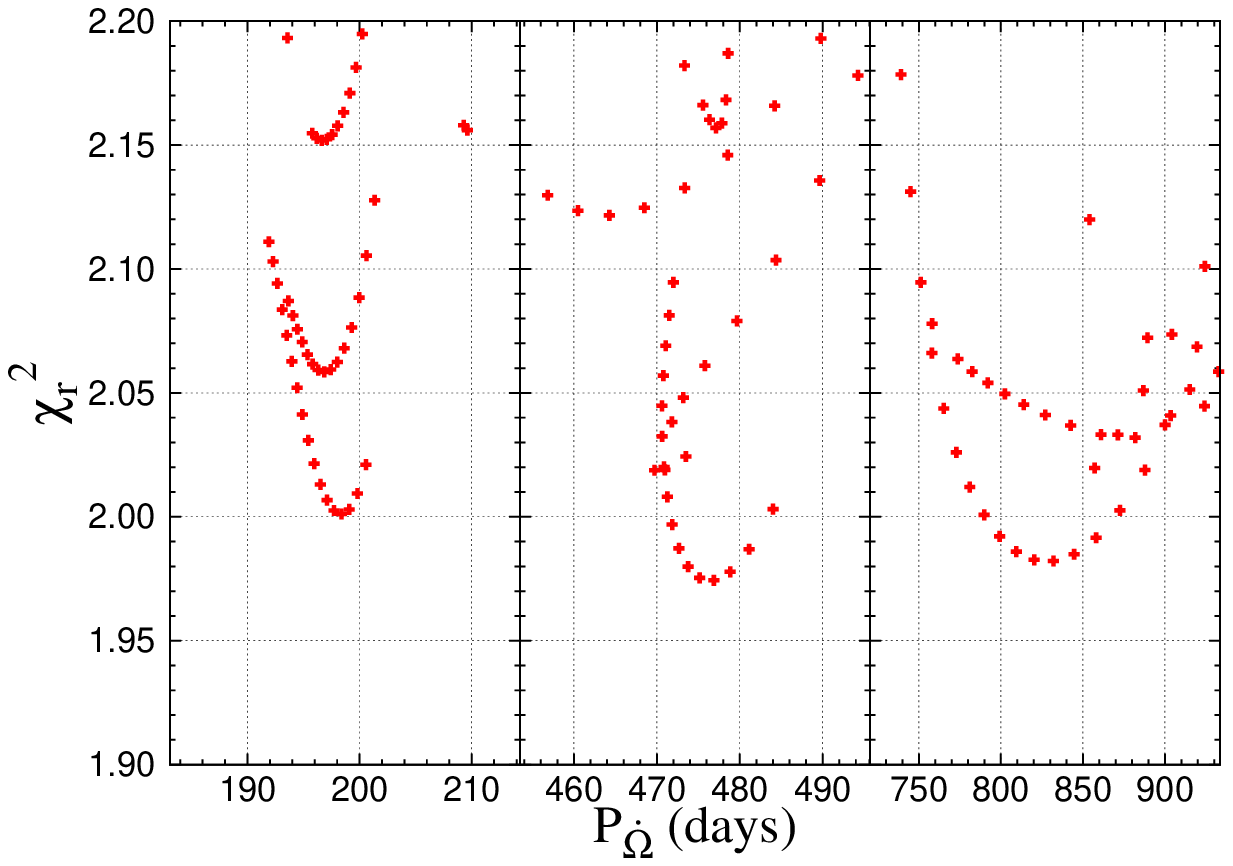}
}%
\vspace{0mm}
\caption{Precession period dependence of $\chi_{r}^2$.
Red crosses are the data points on $P_{\rm{s}}$-$M_{\rm{p}}/M_{\rm{s}}$ plane in Figure \ref{contour} arranged in terms of precession period.
The lower panels are the zoom-ins around $P_{\dot{\Omega}}$=199, 475, and 837 days in the upper panel.}
\label{period-chi}
\end{figure}

It is difficult to see from Figure \ref{contour}, but the resulting solutions are roughly divided into three groups with different precession periods ($P_{\dot{\Omega}} = 2{\pi}/\dot{\Omega}$).
In order to clarify the point, we plot $\chi_{r}^2$ against $P_{\dot{\Omega}}$ in Figure \ref{period-chi}, instead of $P_{\rm{s}}$ or $M_{\rm{p}}/M_{\rm{s}}$.
Now one can easily recognize that $\chi_{r}^2$ is very sensitive to $P_{\dot{\Omega}}$, and the acceptable solutions are located around $P_{\dot{\Omega}} = 199$, 475, and 827 days (see the lower panel in Figure \ref{period-chi}).
For simplicity, we refer to them as short, middle, and long solutions, respectively, in the following discussion.
The best-fit parameters for the three solutions are summarized in Table \ref{solutions} together with those of B13.
Table \ref{solutions} indicates that our solutions prefer the more rapid stellar spin and massive planet regime than those of B13.

\begin{table*}[!htbp]
\centering
\caption{Best fit parameters for our three solutions and those in \citet{2013ApJ...774...53B}.}
\begin{tabular}{cccccc}
\toprule
parameter & short solution & middle solution & long solution & B13 ($M_{\rm{s}}$=0.34$M_{\odot}$) & B13 ($M_{\rm{s}}$=0.44$M_{\odot}$) \\
\midrule
$\rho_{\rm{s}}(\rm{g/cm^3})$ & $0.32\pm0.01$ & $0.27\pm0.01$ & $0.30\pm0.01$ & $0.43\pm0.01$ & $0.57\pm0.02$ \\
$P_{\rm{s}}$(days) & $0.390\pm0.008$ & $0.367\pm0.006$ & $0.331\pm0.008$ & 0.448410 & 0.448413 \\
$i_{\rm{s}} (^{\circ})$ & $123.9\pm2.2$ & $128.2\pm1.7$ & $131.3\pm1.5$ & $119.4\pm0.3$ & $120.3\pm1.3$ \\
$M_{\rm{p}}/M_{\rm{s}}$ & $0.0129\pm0.0014$ & $0.0178\pm0.0009$ & $0.0199\pm0.0008$ & $0.0084\pm0.0006$ & $0.0078\pm0.0007$ \\
$R_{\rm{p}}/R_{\rm{s}}$ & $0.169\pm0.003$ & $0.208\pm0.008$ & $0.289\pm0.016$ & $0.159\pm0.007$ & $0.164\pm0.007$ \\
$i_{\rm{orb}} (^{\circ})$ & $62.9\pm0.8$ & $55.3\pm0.5$ & $50.9\pm0.6$ & $65.2\pm1.6$ & $69.3\pm1.3$ \\
$\Omega (^{\circ})$ & $133.4\pm3.3$ & $131.2\pm2.4$ & $130.2\pm2.2$ & $136.1\pm5.2$ & $125.5\pm0.5$ \\
\midrule
$\phi(^{\circ})$ & $75.3\pm2.5$ & $85.8\pm1.8$ & $92.3\pm1.6$ & $69\pm3$ & $73.1\pm0.6$ \\
$P_{\dot{\Omega}}$(days) & $198.7\pm15.6$ & $474.6\pm21.1$ & $826.9\pm53.3$ & 292.6 & 581.2 \\
\bottomrule
\end{tabular}
\\
\vspace{2mm}
NOTE. - Our angular parameters ($i_{\rm{s}}, i_{\rm{orb}}, \Omega$) are evaluated at $t_{c}$=2455543.9402 HJD while those of B13 are at $t_{c}$=2455536.7680 HJD, whose interval corresponds to 7.1722 days or 16 transits.
Spin-orbit angle $\phi$ is time-invariant.
\label{solutions}
\end{table*}%

\begin{figure}[!h]
\centering
\subfigure{
\includegraphics[width=90mm]{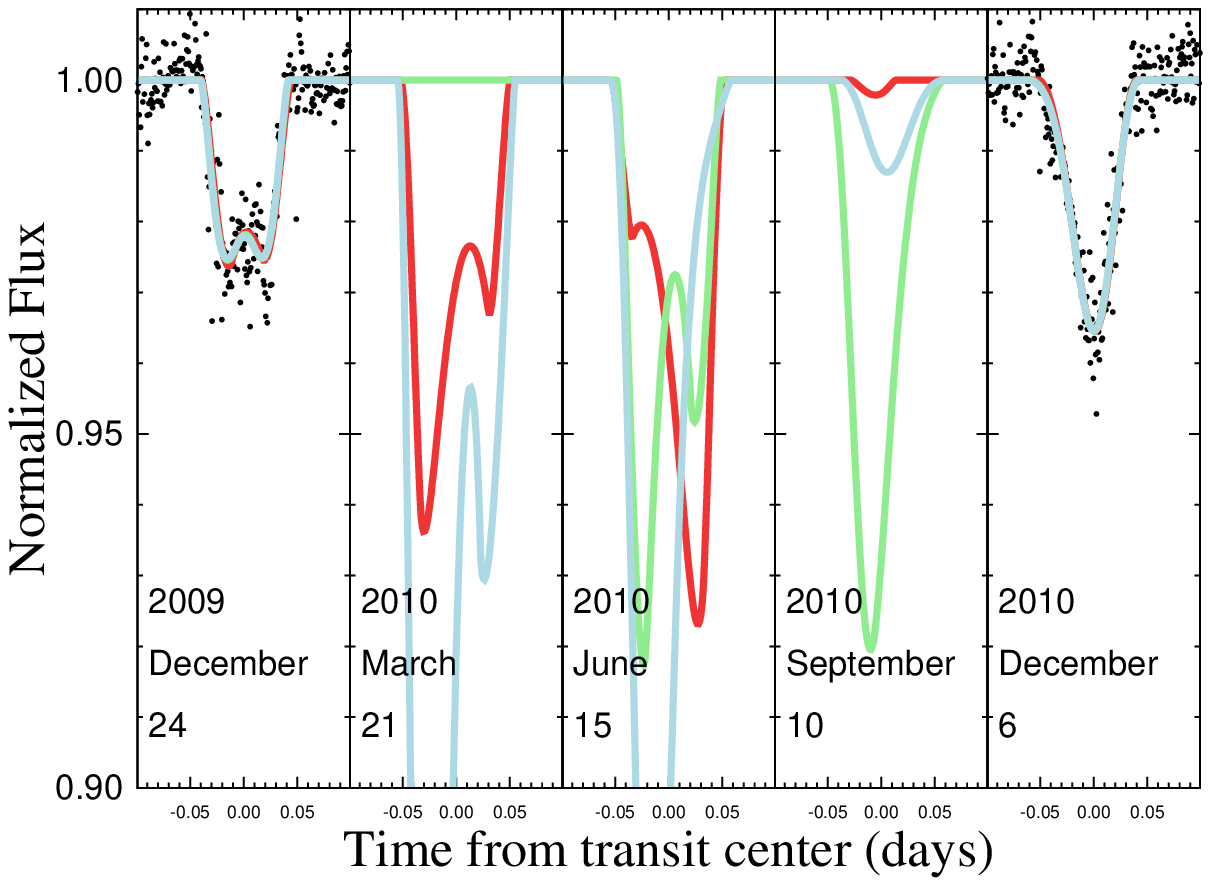}
}%
\\
\vspace{-3mm}
\subfigure{
\includegraphics[width=88mm]{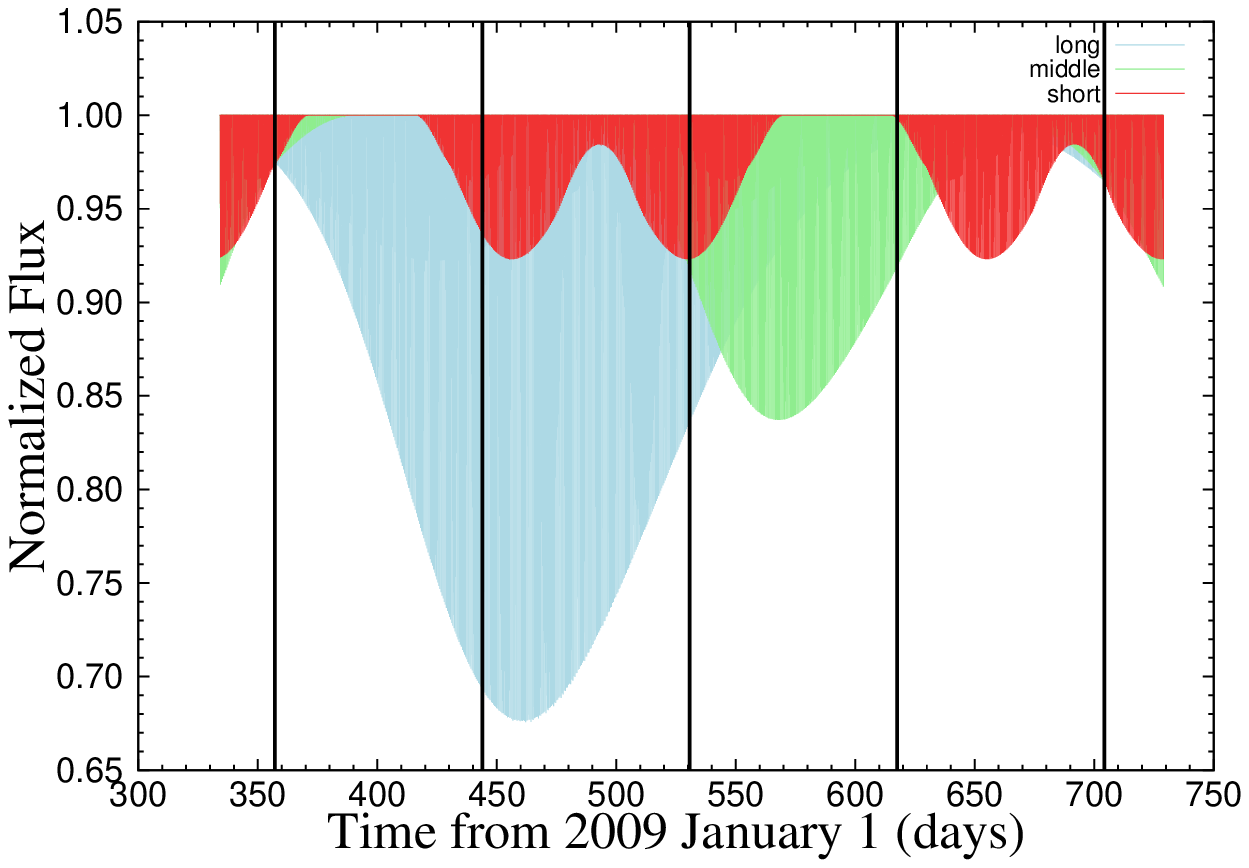}
}%
\vspace{0mm}
\caption{The evolution of the transit lightcurves for three possible solutions in Table \ref{solutions} from 2009 to 2010 observational epochs.
Short, middle and long solutions correspond to red, green, and blue lines, respectively.
Lower panel illustrates the whole evolution from 2009 (leftmost vertical line) to 2010 (rightmost vertical line).
The five vertical lines are drawn for about every three months, each corresponding to the enlarged graph at the upper panel.
Enlarged graphs for 2009 (leftmost) and 2010 (rightmost) are with the observational data (black points), while the intermediate three are only theoretical calculation of lightcurves.}
\label{LC}
\end{figure}

The reason why we have three different solutions can be understood from Figure \ref{LC}, which plots the predicted lightcurves for the three solutions; the short, middle, and long solutions are plotted as red, green, and blue lines, respectively.
Because the transit period is very short, the lightcurves in the lower panel basically show the evolution of the transit $\emph{depth}$ for three solutions.
The five vertical lines in the lower panel are drawn at three-month intervals, each corresponding to the enlarged graph at the upper panels.
By construction, all the three solutions reproduce the 2009 and 2010 data very well (upper-left and upper-right panels), and the significant difference of the lightcurves shows up only during the unobserved epochs.

The examples plotted in the three upper-middle panels illustrate the importance of frequent monitoring of the system in order to specify the system parameters precisely.
Indeed some phases without any transit signals are predicted during their whole evolution because planetary orbital inclination significantly deviates from $90^{\circ}$.
Thus measurements of the transit depth, even with a relatively low signal-to-noise ratio, greatly help distinguishing among the three solution groups.

\subsection{Preliminary comparison with photometry at Koyama Astronomical Observatory
\label{subsec:KAO}}

In this subsection, we attempt a preliminary comparison with the transit photometry taken with Araki telescope at Koyama Astronomical Observatory (KAO) on 2014 November 23, 27, December 2, 23, and 2015 January 10.
The photometry has been performed by a dual-band imager, ADLER (Araki telescope DuaL-band imagER, $2048 \times 2048$ pixels and 12 ${\rm{arcmin}}^2$ FOV for each CCD camera).
The observation was performed with a 670 nm-split dichroic mirror, and the SDSS g$^{\prime}$- and i$^{\prime}$-band filters.
The exposure time is set to 180 s, and 2 exposures are stacked.
The observational data are reduced with standard procedures for photometry, i.e., dark-subtraction, flat-fielding, and aperture photometry ({\rm{daophot}} task in IRAF).
For relative photometry, we adopt a star located at (RA, Dec) = (05:25:07.89, +01:31:16.1) as a reference star.
We also compared the results with other reference stars, and confirmed that the lightcurve did not change significantly, less than $1 \%$.
Since g$^{\prime}$-band data do not have enough signal for the current purpose, we only focus on i$^{\prime}$-band data in this paper.

\begin{figure}[!h]
\centering
\subfigure{
\includegraphics[width=90mm]{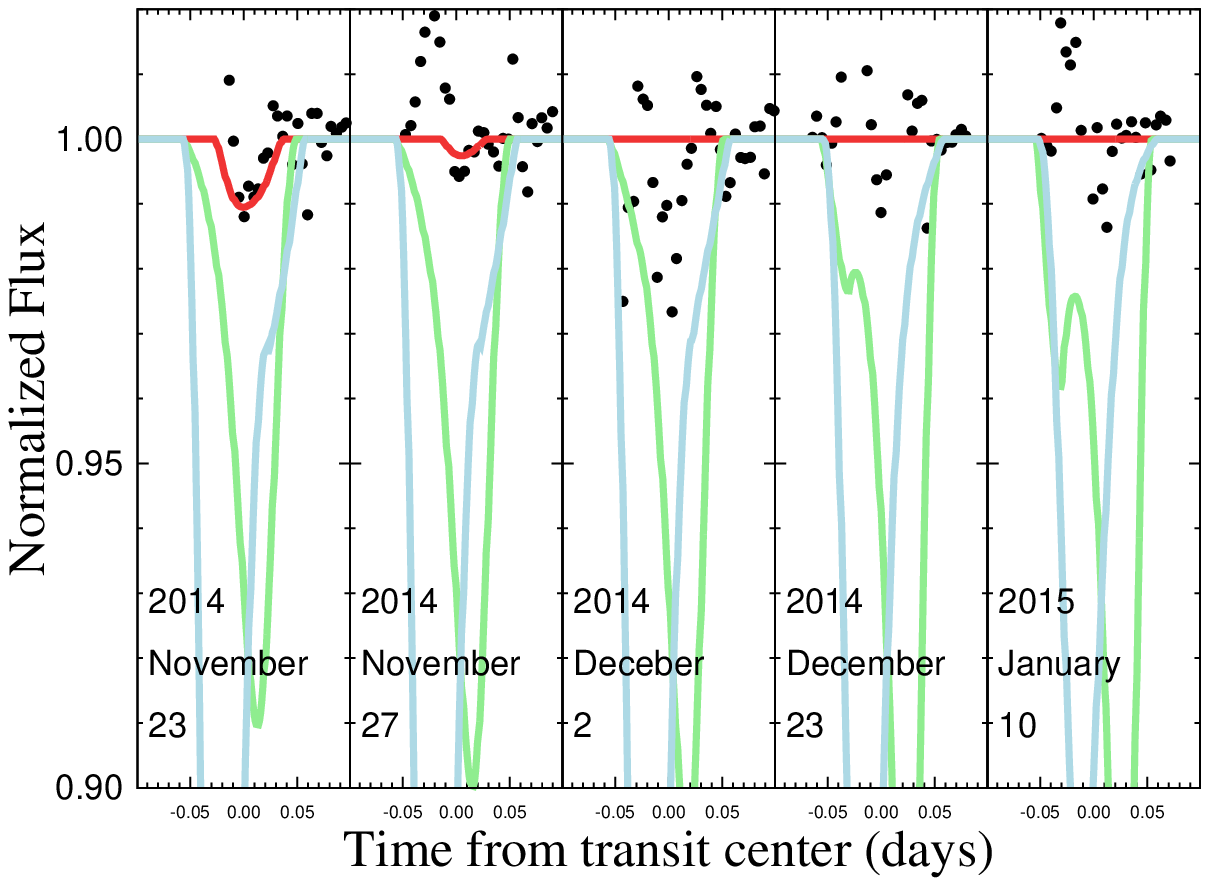}
}%
\\
\vspace{-3mm}
\subfigure{
\includegraphics[width=88mm]{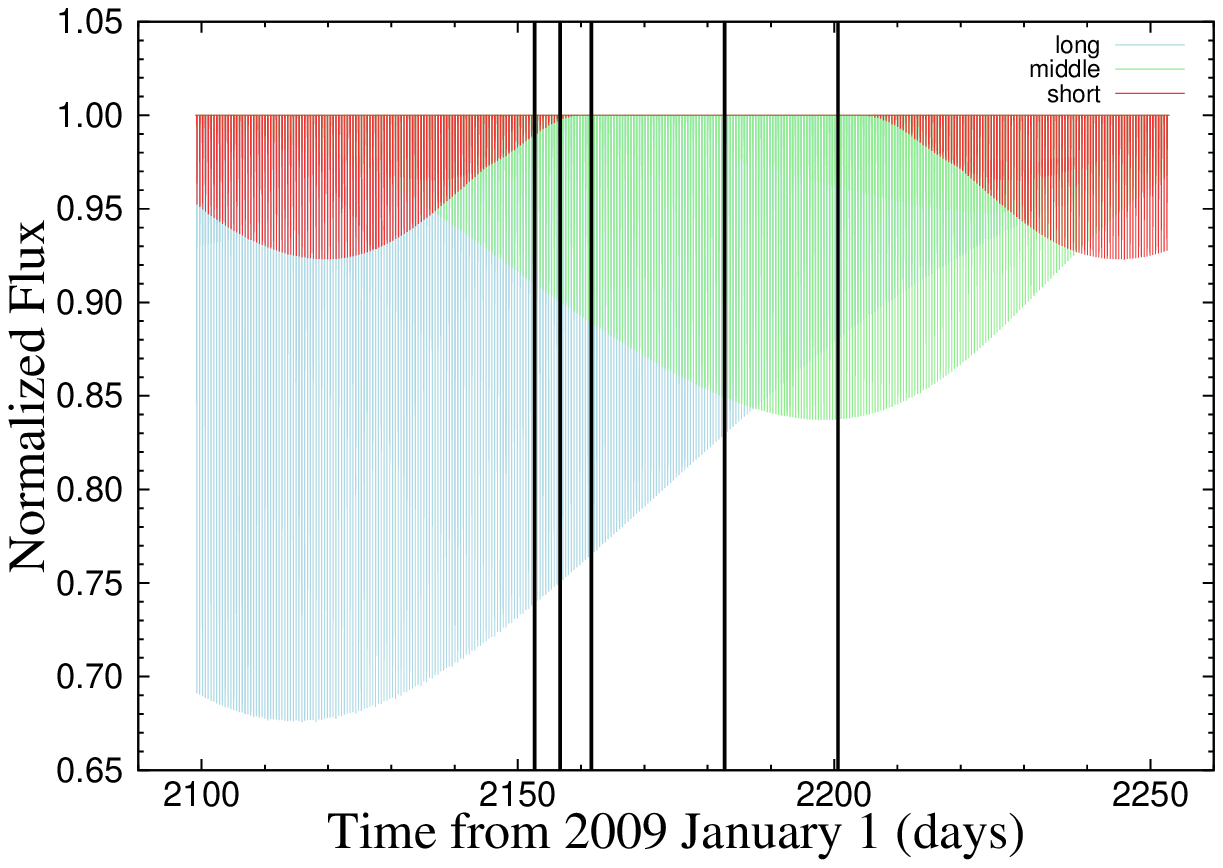}
}%
\vspace{0mm}
\caption{Same as Figure \ref{LC}, but for the observational term of KAO from 2014 November to 2015 January.}
\label{LC_KAO}
\end{figure}

The bottom panel in Figure \ref{LC_KAO} shows the predicted evolution of transit depth for the three solutions over the observed epoch.
The five vertical black lines indicate the above five days, whose enlarged views are shown in the upper panel.
The predicted transit depths at those dates are less than 1$\%$ for the short solution and greater than approximately 10\% for the middle and long solutions.
The filled circles, on the other hand, indicate the data points after removing the long-term stellar activity.
Since there are no clear transit signals beyond the photometric noise level of a few percents, the data seem to prefer the short solution.
Although this conclusion is admittedly very premature, it clearly indicates that such fairly crude photometry may be very useful in distinguishing among the three solutions.
We plan to conduct the photometric observations again during winter in 2015 and 2016.

\section{Long-term dynamical stability \label{sec:tides}}

The results and discussion so far are all focused on the short-term dynamics of the system; the nodal precession whose expected time scale is several hundred days.
In addition to such a short-term behavior, long-term dynamics also contributes to getting an insight into the nature of the pre-main-sequence star and close-in planet system.
When pursuing the longer-term evolution, tidal effect between the star and close-in planet should be considered.
In the PTFO 8-8695 system, particularly, the star and planet are expected to be strongly influenced by the tidal effects because the planetary semi-major axis is very small ($a\leq2R_{\rm{s}}$).

\begin{figure}[!h]
\centering
\includegraphics[width=88mm]{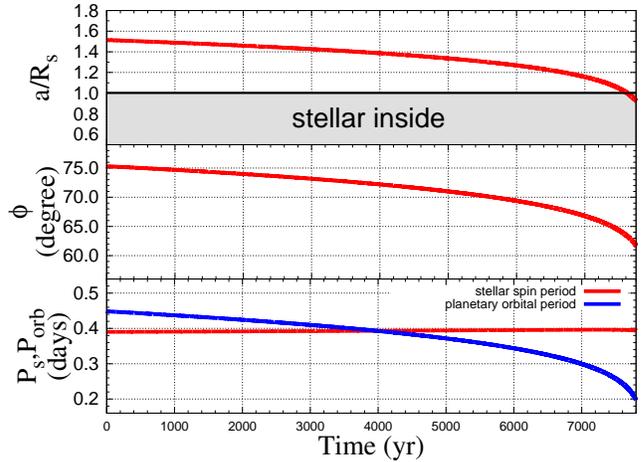}
\caption{Tidal evolution of the planetary orbital semi-major axis (top),
spin-orbit angle $\phi$ (middle), and stellar rotational and planetary
orbital period (bottom).
In the upper panel, stellar inside ($a/R_{\rm{s}}\leq1.0$) is described.
In the lower panel, stellar rotational and planetary orbital periods are shown as red and
blue lines, respectively.
This calculation employs the Love number of $k_{2}=0.028$ and tidal delay time of ${\Delta}t=0.1$ s, both of which are estimated for Sun-like stars.}
\label{tides}
\end{figure}

We investigate the long-term stability of this system following the equilibrium tidal model by \citet{2011CeMDA.111..105C}.
Note here that the two new parameters show up to describe the efficiency of the tidal effect; Love number $k_{2}$ and tidal delay time ${\Delta}t$, both of which refer to the fluid property of the central star.  
Tidal evolution of the system is computed by numerical integration of equations (19) - (21) in \citet{2011CeMDA.111..105C}.
The efficiency of tidal effect is also represented by the reduced tidal quality factor $Q$, which is written in terms of the above parameters as $Q\:{\equiv}\:\frac{3}{4n{k_{2}}{\Delta}t}$ (\citealt{2009MNRAS.395.2268B}, \citealt{2012MNRAS.423..486L}, \citealt{2013ApJ...769L..10R}, \citealt{2014ApJ...784...66X}).
We adopt our short solution as initial conditions and the values of Love number and tidal delay time estimated for Sun-like stars ($k_{2}$ = 0.028 and ${\Delta}t$ = 0.1 s following \citealt{2011CeMDA.111..105C}, which gives $Q\:{\sim}\:10^6$ for an orbital period of 0.448413 days).
The upper, middle, and lower panels of Figure \ref{tides} show the resulting evolution of planetary semi-major axis, spin-orbit angle, and stellar spin and planetary orbital periods, respectively.
Among them, it is worth noting that tidal effect leads to the planetary orbital decay ($a\rightarrow0$) with the time scale less than $10^4$ yrs (upper panel).
This time scale is smaller than the age of PTFO 8-8695 system (2.6-2.7 Myrs) by at least two orders of magnitude.

This rapid orbital decay is totally inconsistent with the observational picture that favors the survival of a close-in planet in the PTFO 8-8695 system.
In addition, both spin-orbit angle $\phi$ and orbital period $P_{\rm{orb}}$ evolve quickly in accordance with the rapid orbital decay (middle panel and blue line in lower panel), which is also incompatible with the significant spin-orbit misalignment or spin-orbit near synchronization obtained by the lightcurve analysis.
Since $k_{2}=0.028$ and ${\Delta}t=0.1$ s are estimated for Sun-like stars, these issues suggest that the internal structure or fluid property of PTFO 8-8695 is significantly different from those of the main-sequence stars.
Reasonable constraints on $k_{2}$ and ${\Delta}t$ of PTFO 8-8695, therefore, allow us to approach the internal structure and fluid properties of pre-main-sequence stars, which are poorly known today due to the scarcity of the observed samples.

\section{Summary and discussion \label{sec:s_and_d}}

The spin-orbit angle $\phi$ plays a key role in investigating the formation mechanism of hot Jupiters (e.g., \citealt{2014A&A...567A..42C} and references therein), and then various methods to determine the angular configuration of the planetary system are developed and utilized.
The most popular one is to use the Rossiter-McLaughlin (RM) effect (\citealt{1924ApJ....60...15R}, \citealt{1924ApJ....60...22M}), which reveals the sky-projected spin-orbit angle through spectroscopy (e.g., \citealt{2005ApJ...622.1118O}, \citealt{2012ApJ...759L..36H}).
By combining the RM effect and asteroseismology, furthermore, three-dimensional spin-orbit angle $\phi$ can be determined (\citealt{2014A&A...570A..54L}, \citealt{2014PASJ...66...94B}).

The stellar gravity-darkening effect provides another methodology to measure $\phi$ through the analysis of photometric transit lightuchrves for rapid rotators (\citealt{2011ApJS..197...10B}, \citealt{2013ApJ...776L..35Z}, \citealt{2014ApJ...786..131A}, \citealt{2015ApJ...805...28M}).
Advantages of this method are that one can determine three-dimensional (i.e., not projected) spin-orbit angle directly and that the analysis is possible without the spectroscopical data (i.e., all we need is the photometric transit lightcurves).
When nodal precession occurs, furthermore, more precise constraints on the system are obtained (\citealt{2013ApJ...768..112P}, \citealt{2013ApJ...774...53B}, \citealt{2015ApJ...805...28M}).

Being a rapid rotator, PTFO 8-8695 is therefore an ideal sample to check the model of gravity darkening and nodal precession.
We reanalyse the time-variable transit lightcurves of this planetary system with the above theoretical model, discarding the spin-orbit synchronous condition assumed in the previous analysis, \citet{2013ApJ...774...53B}.
We find three different groups of solutions that reproduce the observed data, whose precession periods are $199\pm16$, $475\pm21$, and $827\pm53$ days.
These solutions are slightly better, or at least equally good compared to those in the previous work.
The three solutions reproduce 2009 and 2010 observational photometry very well, whereas the theoretically-predicted transit lightcurves at unobserved epochs are found to be totally different.
Difference in transit depth is particularly useful to distinguish among them with the frequent monitoring of the system.
We also present a preliminary comparison of the lightcurve prediction with the additional photometry taken with Araki telescope at Koyama Astronomical Observatory, and find that the solution with the precession period of $199\pm16$ days seems to be preferred.
The solution implies $\rho_{\rm{s}}=0.32\pm0.01\:\:\rm{g/cm}^3$, $P_{\rm{s}}=0.390\pm0.008\:\:\rm{days}$, $M_{\rm{p}}/M_{\rm{s}}=0.0129\pm0.0014$, $R_{\rm{p}}/R_{\rm{s}}=0.169\pm0.003$, and $\phi=75.3^{\circ}{\pm}2.5^{\circ}$.
Although this result is still preliminary, more precise constraints on the system configuration are expected to result from our continuous photometric observations during winter in 2015 and 2016.

The long-term analysis taking into account the tidal effect between star and planet in the PTFO 8-8695 system suggests that tidal parameters $k_{2}$ and ${\Delta}t$ for pre-main-sequence stars are incompatible with those for main-sequence stars; otherwise, the orbiting planet is engulfed by the central star on the very short time scale.
This fact implies that the internal structure of young stars is likely to be different from that for matured stars, and requires further investigation.

Indeed, gravity darkening becomes more manifesting as the star rotates more rapidly.
This is in contrast to the RM effect, which becomes harder to measure in the rapid rotators.
Therefore gravity darkening provides a complementary methodology to the RM effect in the measurement of $\phi$.
Since younger stars are known to rotate more rapidly than those matured, gravity darkening is available to unveil the distribution of $\phi$ for the younger systems by the future observations.
Although tidal evolution of $\phi$ causes the initial memory on the formation process of hot Jupiters to be lost (\citealt{2009MNRAS.395.2268B}, \citealt{2012MNRAS.423..486L}, \citealt{2014ApJ...784...66X}), one can minimize this undesirable effect by preferentially selecting younger systems as targets to measure $\phi$.
These aspects suggest that gravity-darkening (combined with nodal precession, if possible) model is very useful to study the formation scenario of close-in planets.

\bigskip

We are grateful to Julian van Eyken, who kindly provided the photometry data of PTFO 8-8695, and to the referee, Jason W. Barnes, for very constructive comments on the manuscript.
We also thank Masahiro Onitsuka for bringing our attention to PTFO 8-8695, and Masahiro Ikoma, Chelsea X. Huang, and Kengo Tomida for fruitful discussion.
We are also grateful to Mizuki Isogai, Akira Arai, Naofumi Fujishiro, and Hideyo Kawakita for their technical support on observations. 
Data analysis was in part carried out on PC cluster at Center for Computational Astrophysics, National Astronomical Observatory of Japan.
S.K. acknowledges the receipt of a travel grant from the Hayakawa Satio Fund, the Astronomical Society of Japan.
K.M. is supported by JSPS (Japan Society for Promotion of Science) Research Fellowships for Young Scientists (No. 26-7182) and by the Leading Graduate Course for Frontiers of Mathematical Sciences and Physics.
Y.S. and A.Y. gratefully acknowledge the support from Grants-in Aid for Scientific Research by JSPS No. 24340035, and for Young Scientists (B) by JSPS No. 25870893, respectively. 

\bibliographystyle{apj}

\end{document}